\documentclass[letterpaper,11pt]{article}
\usepackage[hmargin=1in,vmargin=1in]{geometry}
\usepackage{amsmath}
\usepackage{graphicx}
\usepackage{bm}
\usepackage{epsfig}
\usepackage{amsmath}
\usepackage{amssymb}
\usepackage{xcolor}
\colorlet{linkequation}{blue}
\usepackage[colorlinks]{hyperref}

\newcommand{\be}{\begin{equation}}
\newcommand{\ee}{\end{equation}}
\newcommand{\bea}{\begin{eqnarray}}
\newcommand{\eea}{\end{eqnarray}}
\begin{document}
\title{Super-horizon effects on the gauge invariant effective energy density in $f(R)$ gravity }
\author{ S. Cheraghchi$^1$, F. Shojai$^{1,2}$\\$^1$Department of Physics, University of Tehran,\\ P.O. Box 14395-547, Tehran, Iran.\\$^2$ School of Physics, Institute for Research in Fundamental Sciences (IPM),\\ P.O. Box 19395-5531, Tehran, Iran.\\}
\date{\today}
\maketitle
\begin{abstract}
In this article we study the gauge invariant effective energy momentum tensor for cosmological perturbations in $f(R)$ gravity  during the inflationary epoch. Considering the super-horizon regime, we derive the effective energy density up to the one-loop corrections. It describes the back-reaction effect of the fluctuations on the background space-time.  
  
\end{abstract}
\section{introduction}
According to the inflationary scenario, the fluctuations of a scalar field, called inflaton, along with the metric perturbations, provide the original seeds for the structure formation of our recent universe, see \cite{dodelson}-\cite{mukhanov 92} and references therein. Due to the non-linearity of Einstein field equations, there is a non-vanishing back-reaction effect through which the fluctuations change the  Friedmann-Lemaître-Robertson-Walker (FLRW) background metric. This effect could appear in the power spectrum of the cosmic fluctuations of CMB \cite{mukhanov 92}. In general relativity (GR), considering the sub-horizon and super-horizon regimes, the back-reaction effects have studied in many articles \cite{abramo barandenberger 97,brandenberger 2015,Geshnizjani 2005,Geshnizjani 2002}. Some issues like local observability and causality \cite{green wald 2014} cause the super-horizon modes have been less studied than the sub-horizon modes. In \cite{unruh}, Unruh has claimed that the effect of long wavelength perturbation is invisible for any local observer but in the recent articles of Brandenberger and his collaborators \cite{brandenberger 2018}, they mention how the super-horizon modes could affect the local physics.\\
In GR, during the slow roll inflation, it is shown that the light scalar field fluctuations along with the scalar metric perturbations  lead to a strong infrared enhancement in the gauge invariant energy momentum tensor (EMT) due to the super-horizon modes, but a combination of slow roll parameters can regulate this infrared enhancement \cite{boyanovsky 2005,boyanovsky 2004,boyanovsky 2006}. In this article we would like to consider the effective EMT in a gauge invariant manner up to the one loop quantum correction of the scalar and tensor perturbations in $f(R)$ gravity with an extra scalar field.
Although $f(R)$ gravity can produce an inflationary era in the early universe, inclusion of some additional scalar fields to it, can lead to  non-adiabatic perturbations. These fields are responsible for primordial non-Gaussianity in the density perturbation \cite{wands}. Moreover multifield inflationary models have rich dynamics and according to \cite{Wienberg} there is no particular reason to believe that a single scalar field drives the inflationary expansion of the universe. These models may also be motivated by high-energy physics, such as supersymmetric theories in which there are many superpartners scalar fields playing a role during the inflationary era \cite{gong}.\\
Considering the effects of super-horizon modes on the gauge invariant EMT during the inflationary epoch in $f(R)$ gravity, here we would see that there is a strong infrared behavior which would be regulated using the slow roll parameters such as GR. 
Using the  effective energy density up to the one-loop order of perturbations amplitude, it could be possible to consider the quantum correction of power spectra \cite{boyanovsky 2005}. However, to find the complete corrections of power spectra, it is necessary to compute the sub-horizon contributions of perturbations in the effective EMT which is not discussed here.

The organization of this paper is as follows. In the next section, we briefly explain the gauge invariant effective EMT. In section 3, metric and scalar field perturbations and their contributions in the time-time component of effective EMT, effective energy density, would be studied.
We will show that the strong infrared behavior is obtained  assuming the nearly scale invariance of the scalar field fluctuations and keeping only the quantum corrections up to the order $\mathcal{O}(H^2\kappa^2)$ ($\kappa^2=8\pi G$). It should be noted, however, that this effect depends on the specific form of the modified gravity. Considering the super-horizon effect of scalar perturbation modes, we have derived the effective potential and power spectrum of curvature perturbation for a special model of \\
$f(R)$ gravity in section 4. 
The gravitational waves back-reaction will be considered in section 5. It is shown that the super-Hubble tensorial modes do not have any infrared enhancement in $f(R)$ gravity such as GR and so during the slow roll inflationary epoch, it would be correct if we calculate the effective energy density for an exact de Sitter background space-time.
\section{ Effective Gauge invariant EMT}
In this section, we would like to study the gauge invariant effective EMT for metric perturbations, both scalar and tensor, in the Jordan frame of $f(R)$ gravity.  This tensor is computed up to the second order of perturbations and describes the back-reaction of the scalar and tensor fluctuations on the inflationary background.  From \cite{mukhanov 92, abramo barandenberger 97},  we know that the back-reaction effects of the scalar and tensor fluctuations are decoupled in GR. Here we follow the method of \cite{abramo barandenberger 97} and extend it for $f(R)$ gravity. \\
As a starting point, we know that because of the homogeneity and isotropy of the background metric, the spatial average of perturbations vanishes. Thus the effective EMT consists of  quadratic terms in metric and scalar fluctuations, but it is not an explicit gauge-invariant expression. To build the corresponding gauge-invariant quantity, let us first consider $f(R)$  action in the Jordan frame  along with a scalar field $\phi$ moving in the potential $V(\phi)$ as
\begin{equation}\label{action1}
S= \int{\sqrt{-g}\left(\frac{1}{2\kappa^2}f(R)-\frac{1}{2}g^{\mu\nu}\partial_
\mu\phi\partial_\nu\phi-V(\phi)\right)}d^4x
\end{equation}
Varying this action with respect to the metric, we obtain the equation of motion for metric
\begin{eqnarray}\label{1}
-\frac{1}{2}g_{\mu\nu}f(R)+FR_{\mu\nu}-\nabla_\mu\nabla_\nu F+g_{\mu\nu}\Box F=\kappa^2\left(\nabla_\mu\phi\nabla_\nu\phi-\frac{1}{2}g_{\mu\nu}\nabla^\lambda\nabla_\lambda\phi-Vg_{\mu\nu}\right)
\end{eqnarray}
where $\Box{F}=(1/\sqrt{-g})\partial_\mu(\sqrt{-g}\partial^\mu F)$. To obtain a gauge invariant EMT, we define a new tensor
\begin{eqnarray}\label{2}
H_{\mu\nu}:=-\frac{1}{2}g_{\mu\nu}f(R)+FR_{\mu\nu}-\nabla_\mu\nabla_\nu F+g_{\mu\nu}\Box F-\kappa^2\left(\nabla_\mu\phi\nabla_\nu\phi-\frac{1}{2}g_{\mu\nu}\nabla^\lambda\nabla_\lambda\phi-Vg_{\mu\nu}\right)
\end{eqnarray}
Expanding the above tensor in terms of metric and the scalar field perturbations up to the second order and then by spatial averaging of it, we get
\begin{align}
H_{\mu\nu}(q_0^a)=-\frac{1}{2\kappa^2}\left\langle H_{\mu\nu,ab}\delta q^a\delta q^b\right\rangle
\end{align}
where $q^a$ is a collective variable whose background value and the corresponding perturbed one are represented by $q_0^a$  and $\delta q^a$, respectively.
The right side of the above relation is of the second order if $\delta q^a$ satisfies the linearized perturbed equations of motion.
Moreover it does not depend on the chosen gauge which means that it would be unchanged under the  general coordinate transformation generated by a given vector field $\xi$ \cite{abramo barandenberger 97}
\begin{equation}\label{x}
\tilde{x}^\alpha=e^{\xi^\beta\partial_\beta}x^\alpha
\end{equation}
leading to the following diffeomorphism for variable $q$
\begin{align}\label{coordinate transformation}
q=q_0+\delta q\to e^{-\mathcal{L}\xi}(q_0+\delta q)=q_0+\delta q-\mathcal{L}_\xi q_0-\mathcal{L}_\xi\delta q+\frac{1}{2}\mathcal{L}^2_\xi q_0
\end{align}
It is assumed that $\left\langle\xi\right\rangle=0$ and $\mathcal{L}_\xi$ denotes the Lie derivative with respect to the vector
field $\xi$. It should be noted that the coordinate transformation (\ref{x}) could not cover all coordinate transformations but as it is  mentioned in \cite{abramo barandenberger 97}, this class of diffeomorphisms is wide enough for our purposes. From (\ref{coordinate transformation}), up to linear order of $\delta q$, we would have
\begin{align}
\delta q\to\widetilde{\delta q}=\delta q-\mathcal{L}_\xi q_0\hspace{0.5in}q_0\to\widetilde{q_0}=q_0-\left\langle\mathcal{L}_\xi\delta q\right\rangle+\frac{1}{2}\left\langle\mathcal{L}_\xi^2\delta q\right\rangle
\end{align}
On the other hand for a tensor field which is constructed from the metric and its derivatives, the following relation would be satisfied \cite{15Ab}
\begin{align}\label{diffeo roul}
(e^{-\mathcal{L}_\xi}G)(x)=G\bigg[\frac{\partial}{\partial x},(e^{\mathcal{L}_\xi}g)(x)\bigg]
\end{align}
Using this, one can explicitely show that $H_{\mu\nu}$ is a gauge invariant quantity \cite{abramo barandenberger 97}. Therefore, it is more appropriate to write it in a gauge invariant form. To do this, instead of $\delta q^a$, it is useful to introduce some gauge invariant quantities,  $\delta Q^a$ (like the Bardeen potentials), which are made from the background variables and their linear perturbations.
 So the gauge invariant EMT takes the following form 
 \begin{align}\label{3}
\left\langle\tau_{\mu\nu}(\delta Q^a)\right\rangle=-\frac{1}{2\kappa^2}\left\langle H_{\mu\nu,ab}\delta Q^a\delta Q^b\right\rangle\equiv-\frac{1}{2\kappa^2}\left\langle\delta H_{\mu\nu}^{(2)}\right\rangle
\end{align}
In the next section, we want to calculate this quantity in the Newtonian gauge for a slow roll inflationary universe. It should be noted that in the above expression, only the expectation values of quadratic perturbations have appeared. Thus, it could be said that,  it is actually  one loop quantum correction of the EMT. 
\section{Scalar perturbations}
To find the effective gauge invariant EMT, at first, we should derive the governing equations for the gauge invariant scalar perturbations.  For a spatially flat FLRW background with the scale factor $a(t)$, the generalized Friedmann equations and the Klein-Gordon equation for the scalar field are \cite{living review 2010}
\begin{equation}\label{b1}
3FH^2=\frac{1}{2}(RF-f)-3H\dot F+\kappa^2\left(\frac{1}{2}\dot\phi_0^2+V(\phi_0)\right)
\end{equation}
\begin{equation}\label{b2}
-2F\dot H=\ddot F-H\dot F+\kappa^2\dot\phi_0^2
\end{equation}
\begin{equation}\label{b3}
\ddot\phi_0+3H\dot\phi_0+V_{,\phi}=0
\end{equation}
where $F(R)\equiv\frac{\partial f}{\partial R}$, $H$ is Hubble parameter, $R=6(2H^2+\dot{H})$    is the Ricci scalar, $\phi_0$ is the background scalar field and dot denotes the time derivative. We consider  small perturbations around FLRW metric
\begin{equation}\label{metric}
ds^2=-(1+2\Phi)dt^2-2a(t) \partial_i\beta dtdx^i+a^2(t)(\delta_{ij}-2\Psi\delta_{ij}+2\partial_i\partial_j\gamma+\mathcal{D}_{ij})dx^idx^j.
\end{equation}
where $\Phi$, $\Psi$, $\gamma$ and $\beta$ are scalar perturbations and $\mathcal{D}_{ij}$ is the symmetric, divergence-less and trace-less tensor perturbation of the metric. Here we choose the Newtonian gauge because it is convenient for calculations, it does not leave a residual gauge symmetry and, more importantly, the gauge invariant combinations of the metric perturbations reduce in this gauge to the metric perturbations  $\Phi$ and $\Psi$. In this gauge
\begin{align}\label{newton gauge}
g_{00}=-1-2\Phi,\hspace{0.5in}g_{0i}=0,\hspace{0.5in}g_{ij}=a^2\delta_{ij}(1-2\Psi) 
\end{align}
and the Fourier transforms of the metric and scalar perturbations, $\delta\phi_k$, satisfy in the following equations \cite{cheraghchi 2018}
\begin{equation}\label{eq2}
\begin{split}
&\left\lbrack3\left(H^2-\dot H+3H\frac{\dot F}{F}\right)+\frac{k^2}{a^2}-\frac{\kappa^2\dot\phi_0^2}{F}\right\rbrack\Phi_k+\left\lbrack3\left(H^2+\dot H-H\frac{\dot F}{F}\right)+\frac{k^2}{a^2}\right\rbrack\Psi_k\\
&+3\left(H+\frac{\dot F}{F}\right)\dot\Psi_k+3H\dot\Phi_k
=-\frac{\kappa^2\dot\phi_0}{F}\delta\dot\phi_k-\frac{\kappa^2}{F}V_{,\phi}\delta\phi_k,
\end{split}
\end{equation}
\begin{equation}\label{eq3}
\left(H+2\frac{\dot{F}}{F}\right)\Phi_k+\dot\Phi_k+\left(H-\frac{\dot{F}}{F}\right)\Psi_k+\dot\Psi_k-\frac{\kappa^2\dot\phi_0}{F}\delta\phi_k=0
\end{equation}
\begin{align}\label{eq4}
\begin{split}
&\ddot\Psi_k+\ddot\Phi_k-\left(\frac{\dot F}{F}-3H\right)\dot\Psi_k+3\left(\frac{\dot F}{F}+H\right)\dot\Phi_k\\
&-\left[\frac{\ddot F}{F}+H\frac{\dot F}{F}-\frac{1}{3}\left(6H^2-\frac{k^2}{a^2}\right)\right]\Psi_k\\
&+\left[2\left(2\dot H+H^2+H\frac{\dot F}{F}\right)-\frac{k^2}{3a^2}+\frac{1}{3F}(4\kappa^2\dot\phi _0^2+3H\dot F+6\ddot F)\right]\Phi_k\\
&=\frac{8\kappa^2\dot\phi_0}{3}\dot{\delta\phi_k}-\frac{4}{3}\kappa^2V_{,\phi}\delta\phi_k
\end{split}
\end{align}
\begin{align}\label{eq5}
\begin{split}
&\ddot\Psi_k+\ddot\Phi_k+\left(\frac{\dot F}{F}+5H\right)\dot\Psi_k+\left(3\frac{\dot F}{F}+5H\right)\dot\Phi_k+\left(\frac{k^2}{3a^2}+\frac{R}{3}-3H\frac{\dot F}{F}-\frac{\ddot F}{F}\right)\Psi_k\\
&+\left[\frac{k^2}{3a^2}-\frac{R}{3}+3\frac{\ddot F}{F}+8H^2+4\dot H+9H\frac{\dot F}{F}+\frac{2}{3F}\kappa^2\dot\phi_0^2\right]\Phi_k\\
&=\frac{2}{3}\kappa^2\dot\phi_0\dot{\delta\phi}_k-\frac{4}{3}\kappa^2 V_{,\phi}\delta\phi_k
\end{split}
\end{align}
\begin{equation}\label{eq6}
\begin{split}
\delta{\ddot\phi}_k+3H\delta\dot\phi_k+\left(\frac{k^2}{a^2}+V_{,\phi\phi}\right)\delta\phi_k=\left(2\ddot\phi_0+6H\dot\phi_0\right)\Phi_k+\dot\phi_0\dot\Phi_k+3\dot\phi_0\dot\Psi_k
\end{split}
\end{equation}
in which we have used the relations
\begin{equation}
-\Phi_k+\Psi_k=\frac{\delta F}{F}
\end{equation}
\begin{align}
\delta R=-2\left\lbrack\left(12H^2+6\dot H-\frac{k^2}{a^2}\right)\Phi_k+3H\dot\Phi_k+2\frac{k^2}{a^2}\Psi_k+3\ddot\Psi_k+12H\dot\Psi_k\right\rbrack.
\end{align}
where $k$ is comoving wavenumber. During the inflationary era, the slow roll parameters are defined as \cite{mukhanov 92} 
\begin{align}\label{slow roll parameters}
\epsilon_1\equiv-\frac{\dot{H}}{H^2},\hspace{0.3in}\epsilon_2\equiv\frac{\ddot{\phi_0}}{2H\dot{\phi_0}},\hspace{0.3in}\epsilon_3\equiv\frac{\dot{F}}{2FH}
\end{align}
Keeping  only the leading order terms in slow roll parameters, equations (\ref{eq2})-(\ref{eq6}) are further simplified to
\be \label{eq8'}
\begin{split}
&\left(H^2(3+\epsilon_1+16\epsilon_3)+\frac{k^2}{a^2}\right)\Phi_k+3H\dot{\Phi}_k+\\
&\left(3H^2(1-\epsilon_1-2\epsilon_3)+\frac{k^2}{a^2}\right)\Psi_k+3H(1+2\epsilon_3)\dot{\Psi}_k=-\frac{\kappa^2\dot{\phi}_0}{F}\delta\dot{\phi}_k-\frac{\kappa^2V_{,\phi}}{F}\delta\phi_k
\end{split}
\ee
\be\label{eq7'}
H(1+4\epsilon_3)\Phi_k+\dot{\Phi}_k+H(1-2\epsilon_3)\Psi_k+\dot{\Psi}_k=\frac{\kappa^2\dot{\phi}_0}{F}\delta\phi_k
\ee
\be\label{eq6'}
\begin{split}
&\left(-\frac{2}{3}\frac{k^2}{a^2}-H^2(2+\frac{32}{3}\epsilon_3+\frac{2}{3}\epsilon_1)\right)\Phi_k-2H\dot{\Phi}_k+\\
&\left(-\frac{2}{3}\frac{k^2}{a^2}-H^2(2-2\epsilon_1-4\epsilon_3)\right)\Psi_k-2H(1+2\epsilon_3)\dot{\Psi}=2\kappa^2\dot{\phi_0}\dot{\delta\phi}_k
\end{split}
\ee
\begin{equation} \label{eq9'}
\begin{split}
\delta{\ddot\phi}_k+3H\delta\dot\phi_k+\left(\frac{k^2}{a^2}-3H^2(-\epsilon_1+2\epsilon_2)\right)\delta\phi_k=2H\dot\phi_0\left(3+2\epsilon_2\right)\Phi_k+\dot\phi_0\dot\Phi_k+3\dot\phi_0\dot\Psi_k
\end{split}
\end{equation}
These equations are highly coupled and thus it is hard to deal with them. To go further, we  use the fact that the action (\ref{action1}) is conformally equivalent to Einstein gravity by defining a minimally coupled new scalar field, $\xi$, which is itself coupled to the $\phi$ field. The study of cosmological perturbations could be simpler in this new frame, the so called Einstein frame. The results  in  the  the  Jordan frame are  finally  obtained  by  an inverse conformal transformation. Before applying the conformal transformation, it would be convenient to express (\ref{metric}) in terms of the conformal time, $\eta=\int{dt/a(t)}$, as follow
\begin{equation}\label{metric1}
ds^2=a^2(\eta)\big[-(1+2\Phi)d\eta^2-2\partial_i\beta d\eta dx^i+(\delta_{ij}-2\Psi\delta_{ij}+2\partial_i\partial_j\gamma+\mathcal{D}_{ij})dx^idx^j\big].
\end{equation}
Now let us use the conformal transformation
\be\label{g}
\tilde{g}_{\mu\nu}=Fg_{\mu\nu}(\eta,\vec{x})
\ee
and also set the auxiliary field and its potential as follows
\be\label{xi eq}
\xi=\sqrt{\frac{3}{2\kappa^2}}\ln F
\ee
\be\label{gg}
V(\xi)=\frac{1}{2\kappa^2}\frac{f-RF}{F^2}
\ee
Perturbing the conformal metric, the relationship between the scalar perturbations of metric in two frames is obtained \cite{mukhanov 92}
\be\label{pert in J and E}
\tilde{\Phi}=\Phi+\frac{\partial \ln F^{\frac{1}{2}}}{\partial R}\bigg|_{R_0}\delta R,\hspace{0.3in}\tilde{\Psi}=\Psi-\frac{\partial \ln F^{\frac{1}{2}}}{\partial R}\bigg|_{R_0}\delta R,\hspace{0.3in }\tilde{\gamma}=\gamma,\hspace{0.3in}\tilde{\beta}=\beta
\ee
where $R_0$ is the background Ricci scalar. From above we see
\be\label{o}
\Psi+\Phi=\tilde{\Psi}+\tilde{\Phi}
\ee
This relation is used further in this section. By using (\ref{g})-(\ref{gg}), action  (\ref{action1}) takes the following form in the conformal frame \cite{Qiu 2015} 
\begin{equation}\label{caction}
S=\int{\sqrt{-\tilde{g}}\left(\frac{\tilde{R}}{2\kappa^2} -  e^{-2 \sqrt{2/3}\kappa\xi}U(\phi) -  V(\xi) -  \tfrac{1}{2} e^{-\sqrt{2/3} \kappa \xi}
\tilde{\nabla}_{a}\phi \tilde{\nabla}^{a}\phi -  \tfrac{1}{2} \tilde{\nabla}_{a}\xi\tilde{\nabla}^{a}\xi\right)}d^4x
\end{equation}
which consists two scalar fields coupled minimally to gravity.
Varying the above action with respect to the conformal metric and scalar fields $\xi$ and $\phi$ yields
\be\label{EFeq1}
e^{-\sqrt{\frac{2}{3}}\kappa\xi_0}\tilde{a}^2U_{,\phi_0}+2\tilde{\mathcal{H}}\phi_0'-\sqrt{\frac{2}{3}}\kappa\phi_0'\xi_0'+\phi_0'' =0
\ee
\be\label{EFeq2}
2\tilde{a}^2\sqrt{\frac{2}{3}}\kappa e^{-2\sqrt{\frac{2}{3}}\kappa\xi_0}U-\tilde{a}^2V_{,\xi}-\frac{e^{-\sqrt{\frac{2}{3}}\kappa\xi_0}}{\sqrt{6}}\kappa\phi_0'^2-2\tilde{\mathcal{H}}\xi_0'-\xi_0''=0
\ee
\be\label{EFeq3}
-\frac{1}{2}\tilde{a}^2e^{-2\sqrt{\frac{2}{3}}\kappa\xi_0}U-\frac{1}{2}\tilde{a}^2V+\frac{3}{2\kappa^2}\tilde{\mathcal{H}}^2-\frac{e^{-\sqrt{\frac{2}{3}}\kappa\xi_0}}{4}\phi_0'^2-\frac{1}{4}\xi_0'^2=0
\ee
in which $\tilde{a}=\sqrt{F}a$ and $\tilde{\mathcal{H}}=\frac{\tilde{a}'}{\tilde{a}}$. A prime denotes derivative with respect to the conformal time and zero index indicates the background fields. By taking the time derivative of the last equation and using equations (\ref{EFeq1}) and (\ref{EFeq2}), we obtain another useful relation 
\be\label{Useful eq}
-\tilde{\mathcal{H}}'+\tilde{\mathcal{H}}^2=\frac{1}{2}\kappa^2\left(e^{-\sqrt{\frac{2}{3}}\kappa\xi_0}\phi_0'^2+\xi_0'^2\right)
\ee
Now using the slow roll parameters defined in Einstein frame \cite{Qiu 2015} 
\be
\tilde{\epsilon}_\xi\equiv\frac{\xi'^2}{2\tilde{\mathcal{H}}^2},\hspace{0.5in}\tilde{\epsilon}_\phi\equiv\frac{e^{-\sqrt{\frac{2}{3}}\kappa\xi_0}\phi_0'^2}{2\tilde{\mathcal{H}}^2}\hspace{0.5in}\tilde{\epsilon}_1\equiv 1-\frac{\tilde{\mathcal{H}}'}{\tilde{\mathcal{H}}^2}
\ee
equations (\ref{EFeq1})-(\ref{EFeq3}) can be written as 
\be\label{EFeq1'}
3\tilde{\mathcal{H}}\phi_0'\sim-\tilde{a}^2e^{-\sqrt{\frac{2}{3}}\kappa\xi_0}U_{,\phi_0}
\ee
\be\label{EFeq2'}
3\tilde{\mathcal{H}}\xi_0'\sim2\tilde{a}^2\sqrt{\frac{2}{3}}e^{-2\sqrt{\frac{2}{3}}\kappa\xi_0}\kappa U-\tilde{a}^2V_{,\xi_0}
\ee
\be\label{EFeq3'}
3\tilde{\mathcal{H}}^2\sim \tilde{a}^2\left(e^{-2\sqrt{\frac{2}{3}}\kappa\xi_0}U+V\right)
\ee
and the perturbed Einstein equations in the Newtonian gauge take the following form
\be\label{EPeq1}
\begin{split}
&\left[-3\tilde{a}^2e^{-2\sqrt{\frac{2}{3}}\kappa\xi_0}U-3\tilde{a}^2V+\frac{6\tilde{\mathcal{H}}^2}{\kappa^2}-e^{-\sqrt{\frac{2}{3}}\kappa\xi_0}\phi_0'^2-{\xi'_0}^2+\frac{\nabla^2}{\kappa^2\tilde{a}^2}\right]\tilde{\Psi}-\frac{3\tilde{\mathcal{H}}}{\kappa^2}\tilde{\Psi}'=\\
&\frac{1}{2}e^{-\sqrt{\frac{2}{3}}\kappa\xi_0}\phi_0'\delta\phi'+\frac{1}{2}\xi_0'\delta\xi'+\frac{1}{2}e^{-2\sqrt{\frac{2}{3}}\kappa\xi_0}U_{,\phi}\delta\phi+\left[-\sqrt{\frac{2}{3}}\tilde{a}^2e^{-2\sqrt{\frac{2}{3}}\kappa\xi_0}\kappa U+\frac{1}{2}\tilde{a}^2V_{,\xi}-\frac{1}{2\sqrt{6}}e^{-2\sqrt{\frac{2}{3}}\kappa\xi_0}\kappa\phi_0'^2\right]\delta{\xi}
\end{split}
\ee
\be\label{EPeq2}
\begin{split}
&\left[-3\tilde{a}^2e^{-2\sqrt{\frac{2}{3}}\kappa\xi_0}U-3\tilde{a}^2V+2\left(e^{-\sqrt{\frac{2}{3}}\kappa\xi_0}\phi_0'^2+\xi_0'^2+\frac{2}{\kappa^2}(-\tilde{\mathcal{H}}^2+2\frac{\tilde{a}''}{\tilde{a}})\right)\right]\tilde{\Psi}+\frac{3\tilde{\mathcal{H}}}{\kappa^2}\tilde{\Psi}'+\frac{1}{\kappa^2}\tilde{\Psi}''=\\
&\frac{1}{2}e^{-\sqrt{\frac{2}{3}}\kappa\xi_0}\phi_0'\delta\phi'+\frac{1}{2}\xi_0'\delta\xi'-\frac{1}{2}\tilde{a}^2e^{-2\sqrt{\frac{2}{3}}\kappa\xi_0}U_{,\phi}\delta\phi+\frac{1}{2\sqrt{6}}\left[4\tilde{a}^2e^{-2\sqrt{\frac{2}{3}}\kappa\xi_0}\kappa U-\sqrt{6}\tilde{a}^2V_{,\xi}-e^{-2\sqrt{\frac{2}{3}}\kappa\xi_0}\kappa\phi_0'^2\right]\delta{\xi}
\end{split}
\ee
\be\label{EPeq3}
e^{-\sqrt{\frac{2}{3}}\kappa\xi_0}\kappa^2\phi_0'\delta\phi+\kappa^2\xi_0'\delta\xi=2\tilde{\mathcal{H}}\tilde{\Psi}+2\tilde{\Psi}'
\ee
$\delta\xi$ is the fluctuation of $\xi$ and we have used the fact that $\tilde{\Phi}=\tilde{\Psi}$ in the absence of any anisotropic stress. Subtracting (\ref{EPeq1}) from (\ref{EPeq2}) and using relations (\ref{Useful eq}), (\ref{EFeq1'}), (\ref{EFeq2'}) and (\ref{EPeq3}), after a little algebra give 
\be\label{tilde{psi}}
2(\tilde{\mathcal{H}}^2-\tilde{\mathcal{H}}')\tilde{\Psi}-\tilde{\Psi}''+ \nabla^2\tilde{\Psi}=0
\ee
In the super-horizon regime, it is useful to define a new time variable $\tilde{t}=\int{\tilde{a}(\eta)d\eta}$ and then write the above equation as 
	\be\label{psitildecosmictime}
	\frac{d^2}{d\tilde{t}^2}\tilde{\Psi}+\tilde{H}\frac{d}{d\tilde{t}}\tilde{\Psi}=0
	\ee
where  $\tilde{H}=\frac{1}{\tilde{a}}\frac{d\tilde{a}}{d\tilde{t}}$. Then, the solution of this equation is 
\be
\tilde{\Psi}=C+D \tilde{t}^{1-\frac{1}{\tilde{\epsilon_1}}}
\ee
in which $C$ and $D$ are integration constants. The first term represents the dominated non-zero constant mode, while the second term is the decaying mode which can be ignored. 
Also, from relation (\ref{o}), we find that
\be\label{oo}
\Psi+\Phi=2\tilde{\Psi}\sim 2C
\ee
To simplify further analysis, without loss of generality, we may write this constant as 
\be\label{oo constant}
C=-\frac{h}{2}(\epsilon_1+\epsilon_3)
\ee
where $h$ is another constant which can be obtained by the initial condition of metric perturbation in the Jordan frame at the time of horizon crossing \cite{cheraghchi 2018}.  By using equation (\ref{xi eq}), it is straightforward to see that
\be
\frac{\delta F}{2F}=\sqrt{\frac{\kappa^2}{6}}\delta\xi
\ee
 Substituting the above results into (\ref{EPeq3}) and using the first two relations of (\ref{pert in J and E}), we find that
\be
\frac{\kappa^2}{F}\phi_0'\delta\phi+3\frac{F'}{F}(\tilde{\Psi}-\Phi)=2\tilde{\mathcal{H}}\tilde{\Psi}+2\tilde{\Psi}'
\ee
Now from (\ref{pert in J and E}),  we can find the scalar metric perturbations in the Jordan frame, for super-horizon regime, as following
\begin{align}\label{Phi1}
\Phi=-\frac{2}{3}\frac{F^2}{aF'}\left(\frac{a}{F}\tilde{\Psi}\right)'+\frac{\kappa^2{\phi_0}'}{3F'}\delta\phi\sim \frac{h}{6}\bigg(\frac{\epsilon_1+\epsilon_3}{\epsilon_3}\bigg)+\frac{\mathcal{H}}{3{\phi_0}'}\left(\frac{\epsilon_1+\epsilon_3}{\epsilon_3}\right)\delta\phi+\mathcal{O}(\epsilon)
\end{align} 
\begin{align}\label{Psi1}
\Psi=\frac{2}{3}\frac{1}{aFF'}\left(a F^2\tilde{\Psi}\right)'-\frac{\kappa^2{\phi_0}'}{3F'}\delta\phi\sim \frac{h}{6}\bigg(\frac{\epsilon_1+\epsilon_3}{\epsilon_3}\bigg)-\frac{\mathcal{H}}{3{\phi_0}'}\left(\frac{\epsilon_1+\epsilon_3}{\epsilon_3}\right)\delta\phi+\mathcal{O}(\epsilon)
\end{align} 
where $\mathcal{H}=\frac{a'}{a}$. Also we have used the definition of slow roll parameters (\ref{slow roll parameters}) and the two background equations (\ref{b2}) and (\ref{b3}). Here, one can easily determine the order of magnitude of $C$ introduced in (\ref{oo constant}). In GR, in the super-horizon regime, the curvature perturbation, $\mathcal{R}=\Psi-\frac{\mathcal{H}}{{\phi}'_0}\delta\phi$  is of the order of
$\frac{1}{\sqrt{\epsilon_1}}$, see relation (10.3.16) in \cite{Wienberg}. On the other hand, according to \cite{cheraghchi 2018} and \cite{Hwang1}, in $f(R)$ gravity, the curvature perturbation, in the super-horizon limit, has a dominated constant mode of this order. Hence by comparing this definition of the  curvature perturbation with equation (\ref{Psi1}), we result that the constant $\frac{h}{6}\bigg(\frac{\epsilon_1+\epsilon_3}{\epsilon_3}\bigg)$ is of the order of $\frac{1}{\sqrt{\epsilon_1}}$.

Substituting (\ref{Phi1}) and (\ref{Psi1}) into (\ref{eq9'}) and rearranging yield the dynamical equation of scalar field fluctuation in the Jordan frame\footnote{Note that $\tilde{\Psi}=\tilde{\Phi}=C$, thus according to (\ref{pert in J and E}), $\Psi$ and $\Phi$ are in the same order of slow roll parameters so their time derivatives would be also of the same order. By using equation (\ref{oo}), it is clear that $\dot\Psi+\dot\Phi\sim\mathcal{O}(\epsilon^{3/2})$, therefore 
we can ignore the time derivative of metric perturbations in favor of $\Psi$ or $\Phi$.} 
\begin{align}\label{deltaphi}
\ddot{\delta\phi}_k+3H\dot{\delta\phi}_k+\left(\frac{k^2}{a^2}-3H^2\big(\epsilon_1-2\epsilon_2-\frac{2}{3}\frac{\epsilon_1+\epsilon_3}{\epsilon_3}\big)\right)\delta\phi_k=hH\dot{\phi_0}\bigg(\frac{\epsilon_1+\epsilon_3}{\epsilon_3}\bigg)
\end{align}
Introducing a new variable $\chi_k=a\delta\phi_k$ and converting cosmic time to conformal time, equation (\ref{deltaphi}) now reads
\be\label{chi}
\chi_k''+\bigg(k^2-\frac{\nu_{\delta\phi}^2-\frac{1}{4}}{\eta^2} \bigg)\chi_k=\frac{3}{2}a h (\epsilon_1-\eta_{\delta\phi}-2\epsilon_2)\mathcal{H}\phi'_0
\ee
On the other hand, it is well known that the slow roll approximation corresponds to an expansion around de Sitter space-time with the scale factor $a(\eta)\propto{\mid\eta\mid}^{-1}$. Thus from the definition of the first slow roll parameter, (\ref{slow roll parameters}), one can easily show that 
\begin{align}\label{a with slow roll approx}
a(\eta)=-\frac{1}{H\eta(1-\epsilon_1)}
\end{align}
Substituting (\ref{a with slow roll approx}) in (\ref{chi}), gives 
\begin{align}\label{J}
\chi_k(\eta)=\frac{1}{2}\sqrt{\pi|\eta|}e^{i(\pi/2)(\nu_{\delta\phi}+1/2)}H_{\nu_{\delta\phi}}^{(1)}(k|\eta|)+ P.S.
\end{align}
in which $\nu_{\delta\phi}=\frac{3}{2}+\epsilon_1-\eta_{\delta\phi}+\mathcal{O}(\epsilon^2,\eta_{\delta\phi}^2)$,  $\eta_{\delta\phi}=\epsilon_1-2\epsilon_2-\frac{2}{3}\frac{\epsilon_1+\epsilon_3}{\epsilon_3}$. Also $H_{\nu_{\delta\phi}}^{(1)}(k|\eta|)$ is  the first kind Hankel function of order $\nu_{\delta\phi}$ and $P.S.$ denotes the particular solution of equation (\ref{chi}) which can be written in terms of the Green function as follow
 \be \label{pss}
 P.S.=\frac{3}{2}h(\epsilon_1-\eta_{\delta\phi}-2\epsilon_2)\int_{\eta_*}^{\eta}G(\eta,\eta')a\mathcal{H}\phi'_0 d\eta'
 \ee
where the lower limit $\eta_*=-1/k$ refers to the lower boundary of the region of integration in which the super-horizon approximation is valid. Using the step function, $\theta(\eta-\eta')$, the retarded Green function can be written as \cite{boyanovsky 2016}
 \be
 G(\eta,\eta')=\frac{1}{3}\bigg(\frac{\eta^2}{\eta'}-\frac{\eta'^2}{\eta}\bigg)\theta(\eta-\eta'),\hspace{0.5in}\eta>\eta'
 \ee
Substituting this expression into (\ref{pss}) and retaining the leading term, give
\be \label{psss}
P.S.=\frac{\eta_0}{6\eta}h(\epsilon_1-\eta_{\delta\phi}-2\epsilon_2)\sqrt{\kappa^2\epsilon_1}\left(3\ln(k\eta)+1-(k\eta)^3\right)+\mathcal{O}(\epsilon)
\ee
where we have written the result in terms of the dimensionless scale parameter $a=\frac{\eta_0}{\eta}+\mathcal{O}(\epsilon)$.\\
In what follows, we consider some models with $\eta_{\delta\phi}\sim\mathcal{O}(\epsilon)\ll1$.  As we will see later, the smallness of  $\eta_{\delta\phi}$ can regulate the simple pole appearing in the effective energy density.
In general, this parameter is not necessarily small but we will show that, under some special conditions, it is of the order of the slow roll parameters. To find a specific model of modified gravity with $\eta_{\delta\phi}\ll1$, note that according to the relations
\be
\epsilon_3=\frac{1}{2HF}\frac{d F}{dR}\dot{R},\hspace{0.5in}12H^2\sim R(1+\frac{\epsilon_1}{2})\hspace{0.5in} \dot{R}\sim-24H^3\epsilon_1
\ee
along with the fact that the time derivative of the slow roll parameters are of the second order, it would be easy to conclude that
\be
F\propto R^{-\frac{\epsilon_3}{\epsilon_1}+\frac{\epsilon_3}{2}}
\ee
Thus $f\propto R^{1-\frac{\epsilon_3}{\epsilon_1}+\frac{\epsilon_3}{2}}$. From the definition of $\eta_{\delta\phi}$, when $\frac{\epsilon_3}{\epsilon_1}+1\sim\mathcal{O}(\epsilon)$, one gets $\eta_{\delta\phi}\ll1$. On the other hand in a model such as $f(R)= R+\frac{\alpha}{\tilde{M}^2}R^{2+\frac{\epsilon_3}{2}}$\footnote{$\tilde{M}^{-2}\equiv \kappa^{2+\epsilon_3}$ and $\alpha$ is a dimensionless constant.}, the second term is dominated at the early times and so, this model can be considered as a toy model here. This means that our desired models of modified gravity have a small deviation from Starobinsky model \cite{starobinsky 75}.  The smallness of  $\eta_{\delta\phi}$ is demanded because it enables us to describe the scalar fluctuations as a light field according to (\ref{chi}). In GR, in the inflationary epoch, the power spectrum of a light scalar field  is nearly scale invariant and here, we would like to consider some specific models with this property. The mentioned model is just an example, but it is not the only one. Hereafter, we would restrict ourselves to such models.\\
Now we are ready to find the effective gauge invariant EMT in  $f(R)$ gravity. Fortunately it is enough to calculate just those terms of the effective EMT which are proportional to $\Phi^2$. At first, we have to find the right hand side of equation (\ref{3}). Equation (\ref{2}) gives the second order expansion of $H_{\mu\nu}$ as  
\begin{align}\label{4}
\begin{split}
&\delta H_{\mu\nu,ab}^{(2)}=\left(-\frac{1}{4}\frac{\partial^2f}{\partial R^2}(\delta R^{(1)})^2-\frac{1}{2}\frac{\partial f}{\partial R}\delta R^{(2)}+\Box\left[\frac{\partial F}{\partial R}\delta R^{(2)}+\frac{1}{2}\frac{\partial^2F}{\partial R^2}(\delta R^{(1)})^2\right]\right)g^0_{\mu\nu}\\
&+\left(\frac{\partial F}{\partial R}\delta R^{(2)}+\frac{1}{2}\frac{\partial^2F}{\partial R^2}(\delta  R^{(1)})^2\right)R^0_{\mu\nu}+\left(\frac{\partial F}{\partial R}\delta R_{\mu\nu} -\frac{1}{2}\frac{\partial f}{\partial R}\delta g_{\mu\nu}\right)\delta R^{(1)}
\\
&-\nabla_\mu\nabla_\nu\left(\frac{1}{2}\frac{\partial^2F}{\partial R^2}(\delta R^{(1)})^2+\frac{\partial F}{\partial  R}\delta R^{(2)}\right)+\Box\left(\frac{\partial F}{\partial R}\delta R^{(1)}\right)\delta g_{\mu\nu}-\delta T_{\mu\nu}^{M(2)}
\end{split}
\end{align}
where $\delta T_{\mu\nu}^M$ is the second order expansion of the scalar field EMT. Since we have restricted our calculation to the super-Hubble modes, it would be possible to ignore all spatial gradients in the above expression. Thus the time-time component of the effective gauge invariant EMT tensor (\ref{3}) has the following form 
\be\label{EEMT}
\begin{split}
&\left\langle\tau_{00}\right\rangle=\\
&\frac{1}{a^2\kappa^2}\bigg[\bigg(216\epsilon_1\frac{\mathcal{H}^4}{a^2}\frac{\partial^2f}{\partial R^2}+(-864-5184\epsilon_1)\frac{\mathcal{H}^6}{a^4}\frac{\partial^3f}{\partial R^3}-20736\epsilon_1\frac{\mathcal{H}^8}{a^6}\frac{\partial^4f}{\partial R^4}\bigg)\left\langle\Phi^2\right\rangle+\\
&\bigg(-12\mathcal{H}\frac{\partial f}{\partial R}-72\epsilon_1\frac{\mathcal{H}^3}{a^2}\frac{\partial^2f}{\partial R^2}+(3456+4320\epsilon_1)\frac{\mathcal{H}^5}{a^4}\frac{\partial^3f}{\partial R^3}+20376\epsilon_1\frac{\mathcal{H}^7}{a^6}\frac{\partial^4f}{\partial R^4}\bigg)\left\langle\Phi\Phi'\right\rangle+\\
&\bigg(3\frac{\partial f}{\partial R}+(198-18\epsilon_1)\frac{\mathcal{H}^4}{a^2}\frac{\partial^2f}{\partial R^2}+(1512-216\epsilon_1)\frac{\mathcal{H}^4}{a^4}\frac{\partial^3f}{\partial R^3}-5184\epsilon_1\frac{\mathcal{H}^6}{a^6}\frac{\partial^4f}{\partial R^4}\bigg)\left\langle\Phi'^2\right\rangle+\\
&\bigg((36-36\epsilon_1)\frac{\mathcal{H}^2}{a^2}\frac{\partial^2f}{\partial R^2}+(432+1944\epsilon_1)\frac{\mathcal{H}^4}{a^4}\frac{\partial^3f}{\partial R^3}+10368\epsilon_1\frac{\mathcal{H}^6}{a^6}\frac{\partial^4f}{\partial R^4}\bigg)\left\langle\Phi\Phi''\right\rangle+\\
&\bigg(-108\frac{\mathcal{H}}{a^2}\frac{\partial^2f}{\partial R^2}-(864+216\epsilon_1\frac{\mathcal{H}^3}{a^4}\frac{\partial^3f}{\partial R^3}-5184\epsilon_1\frac{\mathcal{H}^5}{a^6}\frac{\partial^4f}{\partial R^4}\bigg)\left\langle\Phi'\Phi''\right\rangle+\bigg(-72\frac{\mathcal{H}}{a^2}\frac{\partial^2f}{\partial R^2}-216\epsilon_1\frac{\mathcal{H}^3}{a^2}\frac{\partial^3f}{\partial R^3}\bigg)\left\langle\Phi\Phi'''\right\rangle+\\
&\bigg(-\frac{9}{a^2}\frac{\partial^2f}{\partial R^2}-54(1-\epsilon_1)\frac{\mathcal{H}^2}{a^4}\frac{\partial^3f}{\partial R^3}-1296\epsilon_1\frac{\mathcal{H}^8}{a^6}\frac{\partial^4f}{\partial R^4}\bigg)\left\langle\Phi''^2\right\rangle+108\frac{\mathcal{H}}{a^4}\frac{\partial^3f}{\partial R^3}\left\langle\Phi''\Phi'''\right\rangle+\\
&\bigg(\frac{18}{a^2}\frac{\partial^2f}{\partial R^2}+216\frac{\mathcal{H}^2}{a^4}\frac{\partial^3f}{\partial R^3}\bigg)\left\langle\Phi'\Phi'''\right\rangle\bigg]+\frac{1}{2a^2}\left\langle(\delta\phi')^2\right\rangle-\frac{1}{2}V_{,\phi\phi}\left\langle{\delta\phi}^2\right\rangle-2V_{,\phi}\left\langle\Phi{\delta\phi}\right\rangle-\frac{1}{2a^2}\left\langle(\nabla\delta\phi)^2\right\rangle\\
\end{split}
\ee
in which we have used the relations (\ref{oo}) and (\ref{oo constant}) to write $\Psi$ in terms of $\Phi$ and the expectation values are in the Bunch-Davies vacuum state. By using the above relation and neglecting the time derivative of metric perturbations, the effective energy density up to the leading order of slow roll parameter is
\be\label{tau00}
\begin{split}
&\left\langle\tau^0_{0}\right\rangle_{IR}\simeq\frac{864}{\kappa^2} {H}^6\frac{\partial^3f}{\partial R^3}\left\langle\Phi^2\right\rangle+2V_{,\phi}\left\langle\Phi{\delta\phi}\right\rangle+\frac{1}{2}V_{,\phi\phi}\left\langle{\delta\phi}^2\right\rangle=\\
&\bigg[\frac{96}{\kappa^2}\frac{\partial^3f}{\partial R^3}\frac{H^{8}}{\dot{\phi_0^2}}\left(\frac{\epsilon_1+\epsilon_3}{\epsilon_3}\right)^2-2H^2\left(\frac{\epsilon_1+\epsilon_3}{\epsilon_3}\right)-\frac{3}{2}H^2(2\epsilon_2-\epsilon_1)\bigg]\left\langle\delta\phi^2\right\rangle +\frac{24h^2H^6}{\kappa^2}\bigg(\frac{\epsilon_1+\epsilon_3}{\epsilon_3}\bigg)^2\frac{\partial ^3f}{\partial R^3}
\end{split}
\ee
where we have used the Gaussianity of the fluctuations, $\left\langle\Phi\right\rangle=\left\langle\delta\phi\right\rangle=0$, and relation (\ref{Phi1}) to write $\Phi$ in terms of $\delta\phi$. Also the subscript $IR$ indicates the contribution of the super-horizon modes.  
As it is clear, the first term in the first line of (\ref{tau00}) is made up of the scalar metric perturbations and it does not appear in GR. The last term in (\ref{tau00}) comes from the constant part of relation (\ref{Phi1}), and as mentioned before, it could have a significant contribution in the effective energy density. Now let us pay our attention to calculate the expectation value (\ref{tau00}). The quantized scalar field fluctuation can be expanded as  
\begin{align}
\delta\phi(x,\eta)=\frac{1}{a\sqrt{\Omega}}\sum_{\textbf{k}}{e^{i\textbf{k}.\textbf{x}}\bigg[\chi_k(\eta)b_k+\chi^*_k(\eta)b^*_k\bigg]} 
\end{align}
where $\Omega$ is the spatial volume, $b_k$ and $b^*_k$ are the creation and annihilation operators which obey the standard commutator relations, $k$ is the comoving wave vector and $\chi_k$ is given by (\ref{J}). Using Bunch Davies vacuum, we find that\\
\be\label{expvalue}
\begin{split}
&\left\langle\delta\phi^2\right\rangle\simeq\frac{H^2}{8\pi}\int_0^{\Lambda_p} |H^{(1)}_{\nu_{\delta\phi}}(z)|^2
(z)^3\frac{dz}{z}+\frac{H^2 a\sqrt{a}\gamma}{24\pi^{3/2}\sqrt{\eta_0}}e^{i\pi/2(\nu_{\delta\phi}+1/2)}\int_0^{\Lambda_p}{H^{(1)}_{\nu_{\delta\phi}}(z)\big(3\ln(z)+1-z^3\big)z^3\frac{dz}{z}}\\
&\frac{H^2a^3\gamma^2}{72\pi^2\eta_0}\int_0^{\Lambda_p}{\big(3\ln(z)+1-z^3\big)^2z^2dz}
\end{split}
\ee
where $z=k\eta$, $\gamma=h(\epsilon_1-\eta_{\delta\phi}-2\epsilon_2)\sqrt{\kappa^2\epsilon_1}$ and the cutoff $\Lambda_p$ is an indication of ultraviolet divergences which would be removed by re-normalization. The two last integrals are finite in the infrared limit and thus these terms do not lead to any infrared enhancement. Moreover, at the initial time when $a\to0$, these terms tend to zero, thus we can ignore them. 
Here, we discuss models in which $\eta_{\delta\phi}\ll1$, so the scalar field fluctuation is described by a light scalar field, just like GR. Thus, in these models, beside the ultraviolet divergences, we would have an infrared divergence which appears as a simple pole. This is followed from the first term of (\ref{expvalue}). To solve this integral, we have used the asymptotic behavior of the integrand \cite{boyanovsky 2005}
\be
z^3 \, \left|H^{(1)}_{\nu_{\delta\phi}} (z)\right|^2 \buildrel{z \to 0}\over=\left[
\frac{2^{\nu_\delta\phi} \; \Gamma(\nu_{\delta\phi})}{\pi} \right]^2 \; z^{2 \,
	\Delta_{\delta\phi}}
\ee
where $\Delta_{\delta\phi}=3/2-\nu_{\delta\phi}$ and $\Gamma$ is Gamma function. We thus arrive at 
\be
\left\langle\delta\phi^2\right\rangle\simeq\int_0^{\Lambda_p} |H^{(1)}_{\nu_\delta\phi}(z)|^2
(z)^3\frac{dz}{z}=\frac{1}{2\pi}\left[\ln\Lambda_p^2+\Lambda_p^2+\frac{1}{\Delta_{\delta\phi}}+2\gamma-4+\mathcal{O}(\Delta)\right]
\ee 
where $\gamma$ is the Euler-Mascheroni constant. The simple pole appearing above, would be compensated by the slow roll coefficients in the effective gauge invariant energy density and so would be finite. The contribution of the other terms in the effective energy density, after re-normalization, are some finite values of order $\mathcal{O}(H^4\eta_{\delta\phi})$ or $\mathcal{O}(H^4\epsilon)$ and can be ignored in the super-horizon regime. Substituting these into (\ref{tau00}), finally we get     
\begin{align}
\begin{split}
&\left\langle\tau^0_{0}\right\rangle_{IR}=
\frac{3H^2}{(4\pi)^2}\frac{1}{\eta_{\delta\phi}-\epsilon_1}\bigg[108\frac{H^6}{2}\frac{\partial^3f}{\partial R^3}\frac{\eta_{\delta\phi}^2}{\epsilon_1+\epsilon_3}+3H^2\eta_{\delta\phi}+\frac{3}{2}H^2(2\epsilon_2-\epsilon_1)\bigg]+\\
&\frac{54}{\kappa^2}h^2H^6(\epsilon_1-2\epsilon_2-\eta_{\delta\phi})^2\frac{\partial ^3f}{\partial R^3}+\mathcal{O}(\epsilon)
\end{split}
\end{align}
It might look strange that the super-Hubble modes can influence on the background space-time in such a way that an observer could measure their effects  locally. This is in fact the reason that these modes have less been considered in the literatures than the sub-horizon modes. The effects of super-horizon modes are discussed in detail by Brandenberger and and his co-authors \cite{brandenberger 2015, brandenberger 2018} \footnote{ According to \cite{brandenberger 2015}, it is easy to
understand this point in terms of an analogy with the black-hole physics. A distant observer can detect the gravitational effect of a mass which is  absorbed by a black-hole and disappeared. Similar to this, in cosmology, the super-horizon perturbation modes decay quickly but can affect locally.}

As an instance, we consider the inflationary model $f(R)= R+\frac{\alpha}{\tilde{M}^2}R^{2+\epsilon_3/2}$. In this case, the dominate terms of effective energy density are of the zeroth order of slow roll parameters as follows 
\begin{align}\label{eff}
\begin{split}
\left\langle\tau^0_{0}\right\rangle_{IR}=
\frac{3H^2}{(4\pi)^2}\frac{1}{\eta_{\delta\phi}-\epsilon_1}\bigg[3H^2\eta_{\delta\phi}+\frac{3}{2}H^2(2\epsilon_2-\epsilon_1)\bigg]-\frac{9}{2}\alpha h^2H^4\epsilon_3(\epsilon_1-2\epsilon_2-\eta_{\delta\phi})^2+\mathcal{O}(\epsilon)
\end{split}
\end{align}
This can be compared with the corresponding result in GR \cite{boyanovsky 2005}
\be\label{Eemt00 GR}
\left\langle\tau^0_{0}\right\rangle_{IR}=\frac{3H^4}{(4\pi)^2}\bigg(\frac{2\epsilon_2+3\epsilon_1}{2\epsilon_2+2\epsilon_{1}}\bigg)+\mathcal{O}(\epsilon)
\ee
To clarify more, using the definition of $\eta_{\delta\phi}$ in terms of slow roll parameters, one can simplify the first two terms of (\ref{eff}). This gives 
\begin{align}\label{Eemt00}
\begin{split}
\left\langle\tau^0_{0}\right\rangle_{IR}=
\frac{3H^4}{(4\pi)^2}\frac{2\epsilon_2-\frac{7}{3}\epsilon_{1}}{2\epsilon_2-\frac{2}{3}\epsilon_1}+9\alpha h^2H^4\epsilon_3(\epsilon_1-2\epsilon_2-\eta_{\delta\phi})^2+\mathcal{O}(\epsilon)
\end{split}
\end{align}
Since $h^2\propto\epsilon_1^{-3}$ and $\epsilon_{1}\simeq-\epsilon_3$ in the above mentioned model, the second term of (\ref{Eemt00}) is  also of zeroth order in  the slow-roll parameters. This term would be the dominated term in the effective energy density for some models for which the coupling constant $\alpha$ is of $\mathcal{O}(10^{-1})$, e.g. the Starobinsky model in which $\alpha\sim\frac{1}{6}$.
\section{Effective potential and power spectrum of the curvature perturbation}
Noting the effective energy density (\ref{eff}), one can define the following effective potential 
\be
V_{eff}(\phi_0)=V(\phi_0)+\delta V(\phi_0)=V(\phi_0)+\left\langle\tau^0_0\right\rangle_{IR}
\ee
In our model, $f(R)= R+\frac{\alpha}{\tilde{M}^2}R^{2+\frac{\epsilon_3}{2}}$, the relation $3H^2\simeq\kappa^2V(\phi_0)$ is satisfied up to the zero order of slow roll parameters in inflationary era. This leads
\be
V_{eff}(\phi_0)=V(\phi_0)\left[1+\frac{\kappa^2}{(4\pi)^2}\frac{1}{\epsilon_1-\eta_{\delta\phi}}\bigg(3H^2\eta_{\delta\phi}+\frac{3}{2}H^2(2\epsilon_2-\epsilon_1)\bigg)+3\alpha h^2(\kappa^2H^2)\epsilon_3(\epsilon_1-2\epsilon_2-\eta_{\delta\phi})^2\right]
\ee
This expression gives the effect of super-horizon modes which is of order $\kappa^2H^2$. By this potential, the effective Hubble parameter can be defined as
\be
H_{eff}^2\equiv \frac{\kappa^2}{3}V_{eff}= H^2\big(1+\frac{\delta H^2}{H^2}\big)
\ee
in which
\be
\frac{\delta H^2}{H^2}=\frac{\kappa^2}{(4\pi)^2}\frac{1}{\eta_{\delta\phi}-\epsilon_1}\bigg(3H^2\eta_{\delta\phi}+\frac{3}{2}H^2(2\epsilon_2-\epsilon_1)\bigg)+3\alpha h^2(\kappa^2H^2)\epsilon_3(\epsilon_1-2\epsilon_2-\eta_{\delta\phi})^2
\ee
Now one can find the effective power spectrum of the curvature perturbation as follow \cite{living review 2010}
\be
\mathcal{P}_{eff}^{\mathcal{R}}=\frac{1}{4\pi^2 Q_{s,eff}}H_{eff}^2
\ee
where $Q_{s}\equiv\frac{\dot{\phi}_0^2+\frac{3\dot{F}^2}{2\kappa^2F}}{(H+\frac{\dot{F}}{2F})^2}$ and for our model $Q_{s}=\frac{6}{\kappa^2}F\epsilon_{1}^2\simeq 144\alpha H^2\epsilon_1^2
$. The above relation can be written in a simplified form as follows \cite{living review 2010}
\be
\mathcal{P}_{eff}^{\mathcal{R}}=\frac{1}{(24\pi)^2\alpha\epsilon_{1eff}^2}
 \ee
where the effective slow roll parameter is defined as
 \begin{align}\label{effective slow roll parameter}
 \epsilon_{1eff}\equiv-\frac{\dot{H}_{eff}}{H^2_{eff}}
 \end{align} 
Here, it is useful to express some relations between the slow roll parameters in our model 
\be
\epsilon_1+\epsilon_3\simeq-\epsilon_1\epsilon_3\sim\epsilon_1^2\hspace{0.3in}\epsilon_2=\frac{\ddot{\phi}_0}{2H\dot{\phi}_0}\simeq\frac{\ddot{H}}{4\dot{H}H}\hspace{0.3in}\dot{\epsilon_1}=2H(2\epsilon_1\epsilon_2+\epsilon_1^2)\hspace{0.3in}\dot{\epsilon_2}=H(\epsilon_1\epsilon_2-2\epsilon_2^2+\epsilon_5)
\ee
where $\epsilon_5\equiv\frac{\dot{\ddot{\phi}}_0}{2H^2\dot{\phi}_0}\sim\mathcal{O}(\epsilon^2)$. So we would have
 \be
 H^2_{eff}=H^2\bigg(1+\frac{3\kappa^2 H^2}{64\pi^2}\frac{6\epsilon_2-7\epsilon_1}{3\epsilon_2-\epsilon_1}-\frac{4}{3}\alpha (h\kappa H)^2\epsilon_1^3\bigg)
 \ee   
Using equation (\ref{effective slow roll parameter}), the effective power spectrum of the curvature perturbation is obtained
\be
\begin{split}
&\mathcal{P}^{\mathcal{R}}_{eff}=\frac{1}{(24\pi)^2\alpha}\times\\
&4H^4(\epsilon_1-3\epsilon_2)\bigg(\epsilon_1\big(1+H^2(7\zeta-\sigma\epsilon_1^3)\big)-3\big(1+H^2(2\zeta-\sigma\epsilon_1^3)\big)\epsilon_2\bigg)^3\times\\
&\bigg[3H^4\epsilon_1\bigg(\sigma(2\epsilon_1^5-8\epsilon_1^4\epsilon_2-6\epsilon_1^3\epsilon_2^2+36\epsilon_1^2\epsilon_2^3)+\zeta(5\epsilon_1\epsilon_2+30\epsilon^2_2-5\epsilon_5)\bigg)+\\
&2H^2(\epsilon_1-3\epsilon_2)\epsilon_1\bigg(-(\epsilon_1-3\epsilon_2)+2H^2(-7\zeta+\sigma\epsilon_1^3)\epsilon_1+3H^2(4\zeta-2\sigma\epsilon_1^3)\epsilon_2\bigg)\bigg]^{-2}
\end{split}
\ee
where $\zeta=\frac{3\kappa^2}{64\pi^2}$, $\sigma=\frac{4}{3}\alpha (h\kappa)^2$. The leading term of the last equation is 
\be
\mathcal{P}^{\mathcal{R}}_{eff}=\frac{1}{(24\pi)^2\alpha\epsilon_1^2}\frac{(1+2\zeta H^2)^3}{(1+4\zeta H^2)^2}
\ee
By noting that $\zeta H^2\sim \mathcal{O}(\kappa^2H^2)\ll1$, the above result is compatible with what was found in Starobinsky model \cite{living review 2010}.
\section{Tensor perturbations}
In this section we would consider the tensor perturbation. Since we have assumed that the spatial averaging of any perturbation vanishes, we could ignore the coupling of scalar, tensor and vector perturbations and consider each one as a separate Gaussian field \cite{noh hwang 2001}. On the other hand, according to what is mentioned in \cite{brandenberger 2018}, the physical processes that give rise to the gravitational waves and to the scalar perturbations are disconnected. Keeping this in mind, here,  we would like to study the back-reaction of metric tensor perturbation by considering the effective energy density. Even though in the inflationary era, the back-reaction of metric scalar perturbations would be more important than those of gravitational waves, but considering it could be useful. In $f(R)$ gravity, just like GR, we will show that the effective energy density only depends on the derivatives of $\mathcal{D}_{ij}$ and so it will not appear any infrared singularities and we can set the slow roll parameters to zero. Thus for extracting the leading order of the effective energy density, we could calculate it in the exact de Sitter space-time, like what have been done in \cite{Qiu 2015,duff 77,Christensen 80,birrell} for GR. To do this, at first, we will write the spatial average of the effective energy density at the second order of tensor perturbation as follows and then we will focus on the effects of  the super-horizon modes. As regards, in de Sitter inflation, the sub-horizon modes are in their vacuum state, thus their effects vanish upon renormalization \cite{brandenberger 2018}. As it is mentioned in \cite{brandenberger 2018}, these modes could have a locally measurable effect which would not violate causality, so the studying of these modes would be interesting. In this section we would show that the contribution of the super-horizon tensor modes in the effective energy density, depending on the model, could be negative exactly like GR. In the following we would obtain the full form of the time-time component of effective EMT for the gravitational waves. Since the contribution of long wavelength gravitational waves in the energy density is sub-dominated, $\rho^{IR}_{GW}\ll H^2/\kappa^2$, we will consider these fluctuations as the classical tensor fields \cite{brandenberger 2018}. So equation (\ref{4}) takes the following form
\be\label{TP}
\begin{split}
&\left\langle\tau^{GW}_{00}\right\rangle=\frac{1}{4a^2\kappa^2}F\bigg<-\frac{1}{2}\mathcal{D}'_{ij}\mathcal{D}'^{ij}-4\mathcal{H}\mathcal{D}'_{ij}\mathcal{D}^{ij}+2\nabla_k(\mathcal{D}^{ij}\nabla^k\mathcal{D}_{ij})-\frac{1}{2}\nabla^k\mathcal{D}^{ij}\nabla_k\mathcal{D}_{ij}-\nabla^k\mathcal{D}^{ij}\nabla_j\mathcal{D}_{ik}\bigg>\\
&+\frac{\mathcal{H}}{4a^4\kappa^2}\frac{\partial^2f}{\partial R^2}\bigg<-30\mathcal{D}''_{ij}\mathcal{D}'^{ij}-12\mathcal{D}'''_{ij}\mathcal{D}^{ij}-18\mathcal{H}\mathcal{D}'_{ij}\mathcal{D}'^{ij}-12\mathcal{H}\mathcal{D}''_{ij}\mathcal{D}^{ij}+96\mathcal{H}^2\mathcal{D}'_{ij}\mathcal{D}^{ij}\\
&+9\frac{\mathcal{H}'}{\mathcal{H}}\mathcal{D}'_{ij}\mathcal{D}'^{ij}+12\frac{\mathcal{H}'}{\mathcal{H}}\mathcal{D}''_{ij}\mathcal{D}^{ij}
-12\frac{\mathcal{H}''}{\mathcal{H}}\mathcal{D}'_{ij}\mathcal{D}'^{ij}+\frac{4}{\mathcal{H}}\mathcal{D}''^{ij}\nabla_k\nabla^k\mathcal{D}_{ij}+24\mathcal{D}'^{ij}\nabla_k\nabla^k\mathcal{D}_{ij}\\
&-24\mathcal{H}\mathcal{D}^{ij}\nabla_k\nabla^k\mathcal{D}_{ij}-\frac{12\mathcal{H}'}{\mathcal{H}}\mathcal{D}^{ij}\nabla_k\nabla^k\mathcal{D}_{ij}+\frac{6}{\mathcal{H}}\mathcal{D}'^{ij}\nabla_k\nabla^k\mathcal{D}'_{ij}+24\mathcal{D}^{ij}\nabla_k\nabla^k\mathcal{D'}_{ij}\\
&+\frac{4}{\mathcal{H}}\mathcal{D}^{ij}\nabla_k\nabla^k\mathcal{D}''_{ij}+12\mathcal{H}\nabla^j\mathcal{D}^{ik}\nabla_k\mathcal{D}_{ij}+\frac{6\mathcal{H}'}{\mathcal{H}}\nabla^j\mathcal{D}^{ik}\nabla_k\mathcal{D}_{ij}-12\nabla^j\mathcal{D}'^{ik}\nabla_k\mathcal{D}_{ij}-18\mathcal{H}\nabla^k\mathcal{D}^{ij}\nabla_k\mathcal{D}_{ij}\\
&-\frac{9\mathcal{H}'}{\mathcal{H}}\nabla^k\mathcal{D}^{ij}\nabla_k\mathcal{D}_{ij}+42\nabla^k\mathcal{D}'^{ij}\nabla_k\mathcal{D}_{ij}+\frac{8}{\mathcal{H}}\nabla^k\mathcal{D}''^{ij}\nabla_k\mathcal{D}_{ij}+\frac{6}{\mathcal{H}}\nabla^k\mathcal{D}'^{ij}\nabla_k\mathcal{D}'_{ij}\\
&+\frac{8}{\mathcal{H}}\nabla^k\mathcal{D}^{ij}\nabla_j\nabla_l\nabla_k\mathcal{D}^{l}_{i}-\frac{4}{\mathcal{H}}\nabla_k\nabla^k\mathcal{D}^{ij}\nabla_l\nabla^l\mathcal{D}_{ij}+\frac{4}{\mathcal{H}}\nabla^k\mathcal{D}^{ij}\nabla_l\nabla^l\nabla_j\mathcal{D}_{ik}-\frac{14}{\mathcal{H}}\nabla^\alpha\mathcal{D}^{ij}\nabla_l\nabla^l\nabla_\alpha\mathcal{D}_{ij}\\
&-\frac{4}{\mathcal{H}}\mathcal{D}^{ij}\nabla_k\nabla^k\nabla_l\nabla^l\mathcal{D}_{ij}+\frac{4}{\mathcal{H}}\nabla_k\nabla_j\mathcal{D}_{il}\nabla^l\nabla^k\mathcal{D}^{ij}-\frac{6}{\mathcal{H}}\nabla_l\nabla_k\mathcal{D}_{ij}\nabla^l\nabla^k\mathcal{D}^{ij}\bigg>\\
&+\frac{\mathcal{H}^4}{4a^6\kappa^2}\frac{\partial^3 f}{\partial R^3}\bigg<108\mathcal{D}'_{ij}\mathcal{D}'^{ij}+144\mathcal{D}^{ij}\mathcal{D}''_{ij}+432\mathcal{H}\mathcal{D}^{ij}\mathcal{D}'_{ij}-\frac{54\mathcal{H}''}{\mathcal{H}^3}\mathcal{D}'^{ij}\mathcal{D}'_{ij}-\frac{72\mathcal{H}''}{\mathcal{H}^3}\mathcal{D}^{ij}\mathcal{D}''_{ij}\\
&-\frac{216\mathcal{H}''}{\mathcal{H}^2}\mathcal{D}^{ij}\mathcal{D}'_{ij}+72\nabla_j\mathcal{D}_{ik}\nabla^k\mathcal{D}^{ij}-\frac{36\mathcal{H}'}{\mathcal{H}^3}\nabla_j\mathcal{D}_{ik}\nabla^k\mathcal{D}^{ij}-108\nabla_k\mathcal{D}_{ij}\nabla^k\mathcal{D}^{ij}+\frac{54\mathcal{H}''}{\mathcal{H}^3}\nabla_k\mathcal{D}_{ij}\nabla^k\mathcal{D}^{ij}\\
&-144\mathcal{D}^{ij}\nabla_k\nabla^k\mathcal{D}_{ij}+\frac{72\mathcal{H}''}{\mathcal{H}^3}\mathcal{D}^{ij}\nabla_k\nabla^k\mathcal{D}_{ij}\bigg>
\end{split}
\ee
where $\nabla_k$ is the covariant derivative with respect to the flat FLRW metric and so it can be substituted by the ordinary derivative. Using $\nabla^i\mathcal{D}_{ij}=\mathcal{D}^i_i=0$, the above relation would be simplified to 
\be\label{TEMT}
\begin{split}
&\left\langle\tau^{0 GW}_{0}\right\rangle=\frac{1}{a^2\kappa^2}F\bigg<\frac{1}{8}\mathcal{D}'_{ij}\mathcal{D}'^{ij}+\mathcal{H}\mathcal{D}'_{ij}\mathcal{D}^{ij}+\frac{1}{8}\nabla^k\mathcal{D}^{ij}\nabla_k\mathcal{D}_{ij}\bigg>\\
&+\frac{\mathcal{H}}{4a^4\kappa^2}\frac{\partial^2f}{\partial R^2}\bigg<30\mathcal{D}''_{ij}\mathcal{D}'^{ij}+12\mathcal{D}'''_{ij}\mathcal{D}^{ij}+18\mathcal{H}\mathcal{D}'_{ij}\mathcal{D}'^{ij}+12\mathcal{H}\mathcal{D}''_{ij}\mathcal{D}^{ij}-96\mathcal{H}^2\mathcal{D}'_{ij}\mathcal{D}^{ij}\\
&-9\frac{\mathcal{H}'}{\mathcal{H}}\mathcal{D}'_{ij}\mathcal{D}'^{ij}-12\frac{\mathcal{H}'}{\mathcal{H}}\mathcal{D}''_{ij}\mathcal{D}^{ij}
+12\frac{\mathcal{H}''}{\mathcal{H}}\mathcal{D}'_{ij}\mathcal{D}'^{ij}+6\nabla_k\mathcal{D}'^{ij}\nabla^k\mathcal{D}_{ij}-\left(6\mathcal{H}+3\frac{\mathcal{H}'}{\mathcal{H}}\right)\nabla^k\mathcal{D}^{ij}\nabla_k\mathcal{D}^{ij}\bigg>\\
&+\frac{\mathcal{H}^4}{4a^6\kappa^2}\frac{\partial^3 f}{\partial R^3}\bigg< -108\mathcal{D}'_{ij}\mathcal{D}'^{ij}-144\mathcal{D}^{ij}\mathcal{D}''_{ij}-432\mathcal{H}\mathcal{D}^{ij}\mathcal{D}'_{ij}+\frac{54\mathcal{H}''}{\mathcal{H}^3}\mathcal{D}'^{ij}\mathcal{D}'_{ij}+\frac{72\mathcal{H}''}{\mathcal{H}^3}\mathcal{D}^{ij}\mathcal{D}''_{ij}\\
&+\frac{216\mathcal{H}''}{\mathcal{H}^2}\mathcal{D}^{ij}\mathcal{D}'_{ij}-36\nabla_k\mathcal{D}^{ij}\nabla^k\mathcal{D}_{ij}-\frac{18\mathcal{H}''}{\mathcal{H}^3}\nabla_k\mathcal{D}^{ij}\nabla^k\mathcal{D}_{ij}\bigg>
\end{split}
\ee
To find the effective energy density, let us expand the tensor perturbation in terms of two polarization states as follow 
\be\label{Tensorexpansion}
\mathcal{D}_{ij}(t,x)=\sum_{\lambda=+,\times}\int
\frac{d^3k}{(2\pi)^3}[\epsilon_{ij}(\hat{k},\lambda)e^{i\textbf{k}.\textbf{x}}\mathcal{D}^\lambda_k(t)]
\ee
where $\lambda$ shows the polarization states, the indices $i,j=1,2,3$ and the non-zero polarization tensors are
\be
\epsilon_{11}(\hat{k}=z,+\times)=-\epsilon_{22}(\hat{k}=z,+\times)=\mp i\epsilon_{12}(\hat{k}=z,+\times)=\mp i\epsilon_{21}(\hat{k},+\times)=\frac{1}{\sqrt{2}}
\ee
which are satisfied the transverse-traceless condition and also the following normalization condition 
\be
\sum_{i,j}\left\langle\epsilon_{ij}(\hat{k},\lambda)\epsilon_{ij}(\hat{k'},\lambda ')\right\rangle=2\delta_{\lambda\lambda '}\delta^{(3)}(\hat{k}-\hat{k}')
\ee
The tensor perturbation of metric is a conformal and gauge invariant quantity, so we could consider the evolution of tensor perturbation in the Einstein frame and at the end, we would replace the background functions in the Einstein frame with their corresponding functions in the original frame \cite{mukhanov 92}. The governing equation of tensor perturbation is
\be\label{tp}
\mathcal{D}''_{ij}+2\frac{\tilde{a}'}{\tilde{a}}\mathcal{D}'_{ij}-\nabla^2\mathcal{D}_{ij}=0
\ee
In de-Sitter space-time, equation $(\ref{tp})$ is exactly similar to its GR corresponding. This is because of the fact that Ricci scalar and thus $F$ function are constant. Therefore, it would be possible to use the result of GR in this case \cite{brandenberger 2018}. Writing the above equation in the Fourier space leads the following solution for the Fourier modes of the gravitational waves
\be\label{desitter solution of tensor per}
\mathcal{D}_k=k\eta\bigg[C_1\left(-\frac{\cos (k\eta)}{k\eta}-\sin(k\eta)\right)+C_2\left(-\cos (k\eta)+\frac{\sin (k\eta)}{k\eta}\right)\bigg]
\ee
in which $C_1$ and $C_2$ are the integration constants and we have dropped the index $\lambda$. 
In early times of de Sitter expansion, for the super-horizon modes, the above solution reads 
\be\label{de sitter solution in early uni}
\mathcal{D}_k=A_k(1+\frac{1}{2}(k\eta)^2+\mathcal{O}((k\eta)^2))
\ee 
Then using (\ref{Tensorexpansion}), we would have
\be\label{total tensor's energy density}
\left\langle{\tau^0_0}^{GW}\right\rangle=\frac{1}{\pi^2\kappa^2 a^2}\int{\frac{k^3dk}{k}\left\langle\tilde{\tau}^{0GW}_0\right\rangle(k)}
\ee
where $\tilde{\tau}$ is the Fourier transform of the EMT and has the following form
\be\label{TEMT}
\begin{split}
&\left\langle\tilde\tau^{0 GW}_{0}\right\rangle=F\bigg<\frac{1}{8}|\mathcal{D}'^\lambda_k |^2+\mathcal{H}\mathcal{D}'^\lambda_k\mathcal{D}^{*\lambda}_k+\frac{1}{8}k^2|\mathcal{D}^{\lambda}_k|^2\bigg>\\
&+\frac{\mathcal{H}}{4a^2}\frac{\partial^2f}{\partial R^2}\bigg<30\mathcal{D}''^\lambda_{k}\mathcal{D}'^{*\lambda}_k+12\mathcal{D}'''^\lambda_{k}\mathcal{D}^{*\lambda}_k+18\mathcal{H}|\mathcal{D}'^\lambda_{k}|^2+12\mathcal{H}\mathcal{D}''^\lambda_{k}\mathcal{D}^{*\lambda}_k-96\mathcal{H}^2\mathcal{D}'^\lambda_{k}\mathcal{D}^{*\lambda}_k\\
&-9\frac{\mathcal{H}'}{\mathcal{H}}|\mathcal{D}'^\lambda_{k}|^2-12\frac{\mathcal{H}'}{\mathcal{H}}\mathcal{D}''^\lambda_{k}\mathcal{D}^{*\lambda}_k
+12\frac{\mathcal{H}''}{\mathcal{H}}|\mathcal{D}'^\lambda_{k}|^2+6 k^2\mathcal{D}'^{*\lambda}_k\mathcal{D}^\lambda_{k}-\left(6\mathcal{H}+3\frac{\mathcal{H}'}{\mathcal{H}}\right)k^2\mathcal{D}^{*\lambda}_k\mathcal{D}^{\lambda}_k\bigg>\\
&+\frac{\mathcal{H}^4}{4a^4}\frac{\partial^3 f}{\partial R^3}\bigg< -108|\mathcal{D}'^\lambda_{k}|^2-144\mathcal{D}^{*\lambda}_k\mathcal{D}''^\lambda_{k}-432\mathcal{H}\mathcal{D}^{*\lambda}_k\mathcal{D}'^\lambda_{k}+\frac{54\mathcal{H}''}{\mathcal{H}^3}|\mathcal{D}'^{\lambda}_k|^2+\frac{72\mathcal{H}''}{\mathcal{H}^3}\mathcal{D}^{*\lambda}_k\mathcal{D}''^\lambda_{k}\\
&+\frac{216\mathcal{H}''}{\mathcal{H}^2}\mathcal{D}^{*\lambda}_k\mathcal{D}'^\lambda_{k}-36k^2|\mathcal{D}^\lambda_k|^2-\frac{18\mathcal{H}''}{\mathcal{H}^3}k^2|\mathcal{D}^{\lambda}_k|^2\bigg>
\end{split}
\ee
Denoting the initial amplitude of the gravitational wave by $A_k$, one can define the primordial gravitational wave spectrum as
\be
\mathcal{P}_{prim}(k)=\sum_\lambda \frac{k^3}{\pi^2}|A^\lambda_k|^2
\ee
Hence, (\ref{total tensor's energy density}) would be 
\be\label{e-d of tensor}
\left\langle\tau^{0GW}_0\right\rangle(x)=\frac{1}{\kappa^2 a^2}\int_{k_{min}}^{k_{hor}}\frac{dk}{k}\frac{\left\langle\tilde{\tau}^{0GW}_0\right\rangle(k)}{|A_k^\lambda|^2}\mathcal{P}_{prim}(k)
\ee
in which $k_{min}$ and $k_{hor}$ are the infrared and ultraviolet cutoffs.
By assuming that both polarization states have the same amplitude, hereafter we have dropped the index $\lambda$.
Now using the solution (\ref{de sitter solution in early uni}), we could achieve the Fourier transform of the effective energy density (\ref{TEMT}) up to the leading order of $k$. This means that the kinetic terms and the second and higher order of time derivatives of $\mathcal{D}_k$ are negligible. Using $\mathcal{H}=-\frac{1}{\eta}$ for de Sitter space-time, one finds 
\be\label{Fourier transform of energy density of tensor}
\left\langle\tilde{\tau}^{0GW}_0\right\rangle(k)=\bigg[-\frac{7}{8}F-\frac{87\mathcal{H}^2}{4a^2}\frac{\partial^2f}{\partial R^2}-\frac{72\mathcal{H}^4}{4a^4}\frac{\partial^3f}{\partial R^3}\bigg]k^2\left\langle |A_k|^2\right\rangle
\ee
for the super-horizon modes. The primordial power spectrum of the gravitational waves \cite{brandenberger 2018} is given by
\be\label{power spectrum of tensor pert}
\mathcal{P}_{pirm}=A_T(k^*)\left(\frac{k}{k^*}\right)^{n_T}
\ee
where $k^*$ is the pivot scale by which the tensor spectrum is normalized, $A_T(k^*)$ is the amplitude of power spectrum at that scale and $n_T\simeq -2(\epsilon_1+\epsilon_3)$ is the power spectral index.
Now, by substituting (\ref {Fourier transform of energy density of tensor}) and (\ref{power spectrum of tensor pert}) in (\ref{e-d of tensor}), we have found  that for $n_T>-2$, the effective energy density of tensor fluctuation would be 
\be\label{last tensor per}
\left\langle\tau_0^{0GW}\right\rangle_{IR}=\frac{k_{hor}}{\kappa^2 a^2}\frac{A_T}{2+n_T}\bigg[-\frac{7}{8}F-\frac{87\mathcal{H}^2}{4a^2}\frac{\partial^2f}{\partial R^2}-\frac{72\mathcal{H}^4}{4a^4}\frac{\partial^3f}{\partial R^3}\bigg]\bigg[\left(\frac{k_{hor}}{k^*}\right)^{n_T}-\left(\frac{k_{min}}{k_{hor}}\right)^2\left(\frac{k_{min}}{k^*}\right)^{n_T}\bigg]
\ee
where $k_{hor}=aH$. Here, we consider the tensor perturbation of metric in the classical limit where 
\be\label{constraint}
\left\langle\tau_0^{0GW}\right\rangle_{IR}\ll\frac{H^2(t_i)}{\kappa^2}
\ee
where $t_i$ is the initial time. This demand would result an upper bound on the number of e-folds.
In $f(R)$ gravity, the effective energy density would completely depend on our model. Like the previous section, we consider the model $f(R)= R+\frac{\alpha}{\tilde{M}^2}R^{2+\frac{\epsilon_3}{2}}$, thus equation (\ref{last tensor per}) would be
\be
\left\langle\tau_0^{0GW}\right\rangle_{IR}=\frac{k_{hor}}{\kappa^2 a^2}\frac{A_T}{2+n_T}\bigg[-\frac{7}{8}-64.5H^2\kappa^2\bigg]\bigg[\left(\frac{k_{hor}}{k^*}\right)^{n_T}-\left(\frac{k_{min}}{k_{hor}}\right)^2\left(\frac{k_{min}}{k^*}\right)^{n_T}\bigg]
\ee
where in this model $|n_T|\ll 1$. Thus, the effective energy density has a small correction of order $\kappa^2H^2$. The above negative effective energy density of gravitational waves are exactly similar with the corresponding results in GR.
\section{Conclusion}
In this paper, we have considered the effective energy density in $f(R)$ gravity in the slow roll inflation considering the super-horizon regime. Computing the second order perturbations of Einstein tensor, we have obtained the gauge invariant effective energy density which could be interpreted as the back-reaction of perturbations on the background space-time. In GR, it is shown that the nearly scale invariant power spectrum of the quantum inflaton fluctuations could lead an infrared enhancement of the effective energy density. Here, we have found that choosing some special models of $f(R)$ gravity, it would be possible to have a nearly scale invariant power spectrum for scalar perturbations. Moreover we have shown how the scalar perturbations could have an infrared contribution in the effective energy density in $f(R)$ gravity. This infrared enhancement could be described as a quantum correction in the Friedman equation or indeed effective potential. It is an effect which purely raised by the super-horizon modes.
Moreover, it has not appeared any infrared effect in the effective EMT in $f(R)$ gravity, like GR. It does not mean, however, that the super-horizon tensorial modes do not have any effect on the background space-time. We have shown that depending on $f(R)$ model, the effective EMT of the super-horizon modes, could have a negative sign contribution like in GR.   
\section*{Acknowledgments}
F. Shojai is grateful to the University of Tehran for supporting this work under a grant provided by the university research council.

\end{document}